\global\def\draftcontrol{0}

   \def\versionno{ksbh}

\catcode`\@=11

\expandafter\ifx\csname draftcontrol\endcsname\relax\global\def\draftcontrol{0}
\fi

{\count255=\time\divide\count255 by 60
\xdef\hourmin{\number\count255}
\multiply\count255 by-60\advance\count255 by\time
\xdef\hourmin{\hourmin:\ifnum\count255<10 0\fi\the\count255}}
\def\draftdate{\number\month/\number\day/\number\year\ \ \ \hourmin }

\newcommand\makepapertitle{\par
  \begingroup
    \renewcommand\thefootnote{\@fnsymbol\c@footnote}%
    \def\@makefnmark{\rlap{\@textsuperscript{\normalfont\@thefnmark}}}%
    \long\def\@makefntext##1{\parindent 1em\noindent
            \hb@xt@1.8em{%
                \hss\@textsuperscript{\normalfont\@thefnmark}}##1}%
     \newpage
     \global\@topnum\z@   
     \@makepapertitle
     \thispagestyle{empty}\@thanks
  \endgroup
  \setcounter{footnote}{0}%
  \global\let\thanks\relax
  \global\let\makepapertitle\relax
  \global\let\@makepapertitle\relax
  \global\let\@thanks\@empty
  \global\let\@author\@empty
  \global\let\@date\@empty
  \global\let\@title\@empty
  \global\let\title\relax
  \global\let\author\relax
  \global\let\date\relax
  \global\let\and\relax
  \def\version{\let\version\@version\@gobble}
}
\def\@makepapertitle{%
  \newpage
   \ifnum\draftcontrol=1 {}
   \version\versionno
   \vskip 3em%
   \else
   \hfill\hbox to 3cm {\parbox{4cm}{\@pubnum}\hss}%
   \vskip 3em%
   \fi
   \begin{center}%
   \let \footnote \thanks
     {\LARGE {\@title}}%
     \vskip 1.5em%
     {\normalsize
       \lineskip .5em%
       \begin{tabular}[t]{c}%
         \@author
       \end{tabular}\par}%
     \vskip 1.5em%
     {\@bstract}%
     \end{center}%
     \vskip 1.5em
     \@date%
   \par
}

\gdef\@pubnum{}
\def\pubnum#1{%
  \gdef\@pubnum{#1}}

\gdef\@bstract{}
\def\Abstract#1{%
  \gdef\@bstract{%
   \parbox{\textwidth-0pc}{%
   \centerline{\bf Abstract}\penalty1000%
\kern.2cm%
\noindent
\renewcommand\baselinestretch{1.0}%
{#1}}}
}

\def\ps@paper{\let\@mkboth\@gobbletwo%
     \ifnum\draftcontrol=1
    \def\@oddfoot{\hbox to \textwidth{\tiny \versionno \hfil\tiny\draftdate}%
    \hskip -\textwidth \hbox to \textwidth{\hfil\rm\thepage\hfil}}%
     \else\def\@oddfoot{\hbox to \textwidth{\hfil\rm\thepage\hfil}}
     \fi
     \let\@evenfoot\@oddfoot
}

\def\body{\clearpage
          \pagestyle{paper}
    }

\def\@version#1{\ifnum\draftcontrol=1
\typeout{}\typeout{#1}\typeout{}
\vskip3mm\centerline{\hbox{\fbox{\normalsize{\tt DRAFT -- #1 -- }
                   {\draftdate}}}}\vskip3mm
\fi}
\let\version\@version
\long\def\eqlabel#1{\ifnum\draftcontrol=1
                    \tag@false  
                    \tag*{(\theequation) \hbox to -0.2cm{\hspace{0cm}\small{#1}\hss}}
                    \refstepcounter{equation}
                    \edef\@currentlabel{\theequation}
                    \ltx@label{#1}          
                    \else
                    \label{#1}
                    \fi
                    }
\let\st@bibitem\@bibitem
\let\st@lbibitem\@lbibitem
\ifnum\draftcontrol=1
  \def\@bibitem#1{%
    \st@bibitem{#1}\a@@label{#1}\ignorespaces}
  \def\@lbibitem[#1]#2{%
    \st@lbibitem[#1]{#2}\a@@label{#2}\ignorespaces}
  \def\a@@label#1{%
    \gdef\a@lab{\smash{\normalfont\small#1}}
    \ifvmode
      \if@inlabel
        \global\setbox\@labels\hbox{%
          \llap{\a@lab\let\a@lab\relax
                \kern\@totalleftmargin\kern\marginparsep}%
          \box\@labels}%
      \fi
    \fi}
\fi

\documentclass[12pt,letterpaper]{article}

\usepackage{amsmath,amssymb,array,calc,epsfig,rotating,bm}
\usepackage[sort]{cite}
\usepackage{graphicx}
\usepackage{psfrag,verbatim}
\usepackage{xcolor}
\usepackage{caption}
\usepackage{CJK}

\ifnum\draftcontrol=1
\fi

\renewcommand\baselinestretch{1.25}
\renewcommand\section{\@startsection {section}{1}{\z@}%
                                   {-3.5ex \@plus -1ex \@minus -.2ex}%
                                   {2.3ex \@plus.2ex}%
                                   {\normalfont\large\bfseries}}
\renewcommand\subsection{\@startsection{subsection}{2}{\z@}%
                                   {-3.25ex\@plus -1ex \@minus -.2ex}%
                                   {1.5ex \@plus .2ex}%
                                   {\normalfont\normalsize\bfseries}}
\renewcommand\subsubsection{\@startsection{subsubsection}{3}{\z@}%
                                   {-3.25ex\@plus -1ex \@minus -.2ex}%
                                   {1.5ex \@plus .2ex}%
                                   {\normalfont\normalsize\it}}
\renewcommand\paragraph{\@startsection{paragraph}{4}{\z@}%
                                   {-3.25ex\@plus -1ex \@minus -.2ex}%
                                   {1.5ex \@plus .2ex}%
                                   {\normalfont\normalsize\bf}}

\numberwithin{equation}{section}

\def\revise#1       {\raisebox{-0em}{\rule{3pt}{1em}}%
                     \marginpar{\raisebox{.5em}{\vrule width3pt\
                     \vrule width0pt height 0pt depth0.5em
                     \hbox to 0cm{\hspace{0cm}{%
                     \parbox[t]{4em}{\raggedright\footnotesize{#1}}}\hss}}}}

\def\sqr#1#2{{\vcenter{\vbox{\hrule height.#2pt
 \hbox{\vrule width.#2pt height#1pt \kern#1pt
 \vrule width.#2pt}\hrule height.#2pt}}}}

\def\aa1{\phi}
\def\cc1{\psi}

\catcode`\@=12

\begin{document}


\title{\bf Maxwell's equal-area law with several pairs of conjugate variables for RN-AdS black holes}


\author{
Xiong-Ying Guo,\quad Huai-Fan Li$\footnote{Email: huaifan.li@stu.xjtu.edu.cn; huaifan999@sxdtdx.edu.cn(H.-F. Li)}$,\quad Ren Zhao\\[0.4cm]
\it Department of Physics, \\
\it Shanxi Datong University,  Datong 037009, China\\
\it Institute of Theoretical Physics, \\
\it Shanxi Datong University,  Datong 037009, China\\[0.4cm]
}

\Abstract{In this paper, using Maxwell's equal-area law we study the phase transition of charged AdS black holes by choosing different independent conjugate variables. As is well known, the phase transition can be characterized by the state function of the system, the determination of the phase transition point has nothing to do with the choice of independent conjugate variables. To studying the thermodynamic properties of AdS black holes we give the conditions under which the independent conjugate variables are chosen. When the charge of the black hole is invariable, according to the conditions we find that the phase transition is related to the electric potential and the horizon radius of the charged black hole. Keeping the cosmological constant as a fixed parameter, the phase transition of a charged AdS black hole is related to the ratio of the event horizon to cosmological constant of the black hole. This conclusion is of great importance for us to study the critical phenomenon of black holes and to improve the self-consistent geometry theory of black hole thermodynamic.
}

\makepapertitle

\body

\version\versionno
\tableofcontents

\section{Introduction}
The study of the thermodynamic properties of black holes has always been one of the topics of interest to theoretical physicists. In recent years, the study of black hole thermodynamics in AdS and dS space has attracted wide attentions\cite{Mann1207,Mann1301,Mann1208,Cai,Sekiwa,Jing,Ma16,Ma17,Hendi1,Hendi,Zou,Cheng,Banerjee1,Banerjee2,Banerjee3,Dey,
Bhattacharya,Zeng,Dehyadegari,Cai1606,Zhang1,Zhang2,Wang,Wei,Li:2018aax,Li:2018rpk,Miao,Miao2,Dehyadegari2}. Moreover, the cosmological constant $\Lambda$ is used as the thermodynamic pressure for the AdS black hole in the extended phase space, and its conjugate variable has dimension of volume. Calculating the critical components and finding the phase transition of AdS black hole, this phase transition is analogous to the Van der Waals liquid-gas phase transition~\cite{Mann1207,Mann1301,Mann1208,Cai,Dehyadegari2,Wang,Jafarzade,Zhao,Ali,Dayyani,Zangeneh,Zhang3,Sahay,Zeng2,Wei2,Wei3,Zhao2}. Some researches are done by using the conjugate variables $(P,v)$~\cite{Zeng,Dey,Zhao}, some researches are done by the conjugate variation $(T,S)$~\cite{Mann1207,Cai1606,Dey,Wei3}. In the previous works, we had were studied the critical behaviors of black holes according to $(Q,\Phi)$, through the study we find that the second-order phase transition is independent of the choose of conjugate variable, while the critical exponents are dependent on the choice of conjugate variables~\cite{Li1}.

For a normal canonical thermodynamic system, the phase transition point is the state function of the system. The determination of the phase transition point is independent of the choice of independent conjugate variables. Using the Maxwell's equal area law we discuss that the phase transition points of AdS black hole should be the same with different independent conjugate variables~\cite{Li2,Li3,Zhao4,Zhao5,Zhang5,Belhaj,Spallucci}. When the phase transition points obtained by the independent conjugate variable are different from those obtained by other independent conjugate variable, we believe that this set of independent conjugate variable cannot be regarded as the conjugate variable of AdS black hole. The conditions for selecting independent dual state parameters for studying the thermodynamic properties of black holes are given. It is not only helpful for further understanding the properties of entropy, temperature and heat capacity of black holes, but also is of great importance to the improvement of self-consistent geometry theory of black hole thermodynamic.

This paper in organized as follows. In the next section we review the thermodynamics of charged AdS black hole. In section \ref{eff} we study the critical behavior of charged AdS black hole with Maxwell's equal area law according to different conjugate variables $(P,V)$, $(T,S)$ and $(P,v)$,respectively. we also study the critical behavior by the conjugate variables $(Q^2,\Psi)$ and $(Q,\Phi)$ to find the condition that the phase transition occurs. In section \ref{con}, we give conclusion and discussion for the choice of the conjugate variable in the study of thermodynamics of black hole.

\section{Thermodynamics of charged AdS black hole}
\label{re}

The action of Einstein-Maxwell theory in the
background of AdS spacetime is ~\cite{Mann1207}
\begin{equation}
\label{eq1}
I = \frac{1}{16\pi }\int {d^4} x\sqrt { - g} (R - 2\Lambda - F_{\mu \nu }
F^{\mu \nu }).
\end{equation}
To start with we review some basic thermodynamic properties of the spherical
charged AdS black hole. In Schwarzschild-like coordinates the metric read
\begin{equation}
\label{eq2}
ds^2 = - f(r)dt^2 + f^{ - 1}dr^2 + r^2d\Omega _2^2 ,
\end{equation}
Here, $d\Omega _2^2 $ stands for the standard element on $S^2$ and the function
$f$ is given by
\begin{equation}
\label{eq3}
f(r) = 1 - \frac{2M}{r} + \frac{Q^2}{r^2} + \frac{r^2}{l^2},
\end{equation}
The position of the black hole event horizon is determined as a larger root
of $f(r_ + ) = 0$. The parameter $M$ represents the ADM mass of the black
hole and in our set up it is associated with the enthalpy of the system. $Q$
represents the electric charge. Using the Euclidean trick, one can identify
the black hole temperature
\begin{equation}
\label{eq4}
T = \frac{1}{4\pi r_ + }\left( {1 + \frac{3r_ + ^2 }{l^2} - \frac{Q^2}{r_ +
^2 }} \right) = \frac{1}{4\pi r_ + }\left( {1 + 8\pi Pr_ + ^2 -
\frac{Q^2}{r_ + ^2 }} \right),
\end{equation}
and the corresponding entropy
\begin{equation}
\label{eq5}
S = \frac{A}{4},
\quad
A = 4\pi r_ + ^2 .
\end{equation}

Thus, the ADM mass of black hole is enthalpy $H = M$~\cite{Kastor} and is obtained
as~\cite{Dehyadegari}
\begin{equation}
\label{eq6}
M(S,Q,P) = \frac{1}{6\sqrt {\pi S} }(3S + 8PS^2 + 3\pi Q^2).
\end{equation}
The intensive parameters conjugate to $S$, $Q$ and $P$ are defined by $T =
\left( {\frac{\partial M}{\partial S}} \right)_{P,Q} $, ${\Phi = \left(
{\frac{\partial M}{\partial Q}} \right)_{S,P} } $, $ V = \left(
{\frac{\partial M}{\partial P}} \right)_{S,Q} $, where  $T$  is the
temperature, $\Phi = \frac{Q}{r_ + }$ is the electric potential and the thermodynamics volume $V$ is obtained via $V = \frac{4\pi
}{3}r_ + ^3 $. The new variable $\Psi = \left({\frac{\partial M}{\partial Q^2}} \right)_{S,P} $, which conjugate to $Q^2$,  is the inverse of the specific
volume $v = 2r_ + $ in the natural unit where $l_p = 1$~\cite{Mann1207}. Therefore, the
above thermodynamic relation satisfies the first law as:
\begin{equation}
\label{eq7}
dM = TdS + \Psi dQ^2 + VdP,
\quad
dM = TdS + \Phi dQ + VdP.
\end{equation}
Comparing Eq.(\ref{eq7}) with the first law equations of the general thermodynamics system, we find that
$\Psi (\Phi)$ is conjugated to the $Q^2(Q)$, respectively. From scaling argument, we arrive at the Smarr
formula as :
\begin{equation}
\label{eq8}
M = 2TS + 2\Psi Q^2 - VP,
\quad
M = 2TS + \Phi Q - VP.
\end{equation}
In the previous works, the critical phenomena of charged AdS black hole with different choices of the conjugated variables have been studied.~\cite{Mann1207,Dehyadegari,Cai1606,Zhang1,Zhang2}. Taking $Q$ as a constant, we can obtain
the value of critical point as
\begin{equation}
\label{eq9}
T_c = \frac{\sqrt 6 }{18\pi Q},
\quad
r_c = \sqrt 6 Q,
\quad
P_c = \frac{1}{96\pi Q^2}.
\end{equation}
While taking the $l$ as constant, the value of critical point as
\begin{equation}
\label{eq10}
T_c = \frac{2}{\pi l\sqrt 6 },
\quad
r_c = \frac{l}{\sqrt 6 },
\quad
Q_c = \frac{l}{6}.
\end{equation}

\section{The equal-area law of charged AdS black hole in extended phase space}
\label{eff}

When the charge of the AdS black hole is invariant, the state equation of charged AdS
black hole is given by Eq.(\ref{eq4}). The state equation can be write as $f(T,P,V)=0$, which means
the particle number of system is invariant. First, we investigate
the condition that the phase transition occur with the different conjugate variables $(P,V)$ and $(T,S)$
,respectively, by Maxwell's equal area law.

\subsection{The construction of equal-area law in~$P - V$~diagram}
\label{subsubsec:mylabel1}

\begin{figure}[!htbp]
\center{\includegraphics[width=7cm,keepaspectratio]{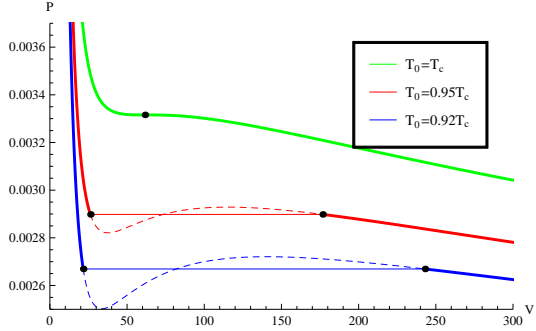}\hspace{0.5cm}}\\
\captionsetup{font={scriptsize}}
\caption{(Color online) The simulated phase transition and the boundary of a two-phase
coexistence on the base of isotherms in the $P-V$ diagram for charged anti-de Sitter black holes.
The temperature of isotherms from top to bottom. The green line is the critical isotherm diagram. In the calculation setting $Q=1$.\label{PVR}}
\end{figure}

When the charge $Q$ of AdS black hole is invariant, the horizontal axis of
two-phase coexistence region in charged AdS black hole are $S_2$ and $S_1$, respectively,
the pressure on the vertical axis as $P_0$, which depends on the horizon radius of
black hole $r_+$. From the Maxwell's equal area
law,
\begin{equation}
\label{eq11}
P_0 (V_2 - V_1 ) = \int\limits_{V_1 }^{V_2 } {PdV},
\end{equation}
we can obtain
\begin{equation}
\label{eq12}
P_0 = \frac{T_0 }{2r_1 } - \frac{1}{8\pi r_1^2 } + \frac{Q^2}{8\pi r_1^4 },
\quad
P_0 = \frac{T_0 }{2r_2 } - \frac{1}{8\pi r_2^2 } + \frac{Q^2}{8\pi r_2^4 },
\end{equation}
\begin{equation}
\label{eq13}
2P_0 = \frac{3T_0 (1 + x)}{2r_2 (1 + x + x^2)} - \frac{3}{4\pi r_2^2 (1 + x
+ x^2)} + \frac{3Q^2}{4\pi r_2^4 x(1 + x + x^2)}.
\end{equation}
here $x = r_1 / r_2 $, $T_0$ is the black hole temperature. From Eq. (\ref{eq12}), we obtain
\begin{equation}
\label{eq14}
0 = T_0 - \frac{1}{4\pi r_2 x}(1 + x) + \frac{Q^2}{4\pi r_2^3 x^3}(1 +
x^2)(1 + x),
\end{equation}
\begin{equation}
\label{eq15}
2P_0 = \frac{T_0 }{2r_2 x}(1 + x) - \frac{1}{8\pi r_2^2 x^2}(1 + x^2) +
\frac{Q^2}{8\pi r_2^4 x^4}(1 + x^4)
\end{equation}
From Eqs.(\ref{eq13}) and (\ref{eq15}), we find that

\begin{equation}
\label{eq16}
\frac{1}{4\pi r_2 x} = \frac{T_0 (1 + x)(1 - 2x + x^2)}{(1 + x - 4x^2 + x^3
+ x^4)} + \frac{Q^2[(1 + x^4)(1 + x + x^2) - 6x^3]}{4\pi r_2^3 x^3(1 + x -
4x^2 + x^3 + x^4)}.
\end{equation}
From Eq. (\ref{eq14}),
\begin{equation}
\label{eq17}
T_0 = \frac{(1 + x)}{4\pi r_2 x} - \frac{Q^2}{4\pi r_2^3 x^3}(1 + x)(1 +
x^2),
\end{equation}
substituting Eq.(\ref{eq17}) into Eq. (\ref{eq16}), we obtain
\begin{equation}
\label{eq18}
r_2^2 = \frac{Q^2}{x^2}\frac{(1 + 2x - 6x^2 + 2x^3 + x^4)}{(1 - x)^2} =
\frac{Q^2}{x^2}(1 + 4x + x^2) = Q^2f_Q (x).
\end{equation}
Substituting Eq.(\ref{eq18}) into Eq. (\ref{eq17}), we find that
\begin{equation}
\label{eq181}
 T_0 = \frac{(1 + x)}{4\pi xQf^{1 / 2}(x)} - \frac{(1 + x)(1 + x^2)}{4\pi Qf^{3 / 2}(x)x^3}
\end{equation}
Taking $T_0 = \chi T_c$, $T_c = \frac{\sqrt 6 }{18\pi Q}$ is critical temperature. When $0 < \chi <
1$, so Eq.(\ref{eq181}) can be rewrite as
\begin{equation}
\label{eq19}
\chi x^2f_Q^{3 / 2} (x)\frac{1}{3\sqrt 6 } = (1 + x).
\end{equation}
We can derive the radius of black hole horizon $r_2 $ or $r_1 $ with Eqs. (\ref{eq18}) and (\ref{eq19}).
We find that the temperature $T_0 $ and the pressure $P_0$ satisfy the conduction of the first-order phase transition according to the Ehrenfest's classification. From Eqs.(\ref{eq4}),(\ref{eq18}) and (3.10), we can plot the phase diagram $P-V$ with
different $\chi $ in Fig.\ref{PVR}.

\subsection{The construction of equal-area law in~$T - S$~diagram}
\label{subsubsec:mylabel2}

\begin{figure}[!htbp]
\center{
\includegraphics[width=7cm,keepaspectratio]{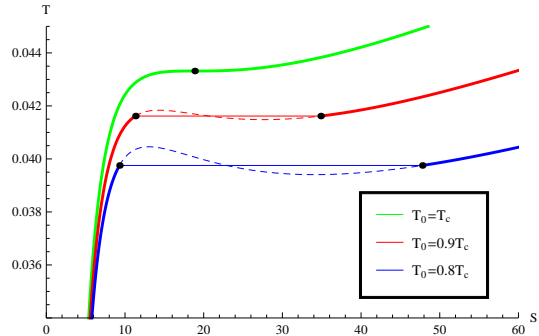}\hspace{0.5cm}}\\
\captionsetup{font={scriptsize}}
\caption{(Color online) The simulated phase transition and the boundary of a two-phase
coexistence on the base of isotherms in the $T-S$ diagram for charged anti-de Sitter black holes.
The temperature of isotherms from top to bottom. The green line is the critical isotherm diagram.\label{TSR}}
\end{figure}

When the cosmological constant $l$ is invariant, the horizontal axis of
two-phase coexistence region in charged AdS black
hole are $S_2$ and $S_1$, respectively, the temperature on the vertical axis as $T_0 (T_0 \le
T_c )$, which depend on the horizon radius of black hole $r_+$. From the Maxwell's equal area
law,
\begin{equation}
\label{eq20}
T_0 (S_2 - S_1 ) = \int\limits_{S_1 }^{S_2 } {TdS} = \int\limits_{r_1
}^{r_2 } {\frac{1}{2}\left( {1 + \frac{3r_ + ^2 }{l^2} - \frac{Q^2}{r_ + ^2
}} \right)dr_ + } ,
\end{equation}
\begin{equation}
\label{eq21}
T_0 = \frac{1}{4\pi r_2 }\left( {1 + \frac{3r_2^2 }{l^2} - \frac{Q_0^2
}{r_2^2 }} \right),
\quad
T_0 = \frac{1}{4\pi r_1 }\left( {1 + \frac{3r_1^2 }{l^2} - \frac{Q_0^2
}{r_1^2 }} \right),
\end{equation}
\begin{equation}
\label{eq22}
2\pi T_0 r_2^2 (1 - x^2) = r_2 (1 - x) + \frac{r_2^3 }{l^2}(1 - x^3) - Q_0^2
\frac{1 - x}{r_2 x}.
\end{equation}
From Eq.(\ref{eq21}), we can obtain
\begin{equation}
\label{eq23}
0 = - \frac{r_2 - r_1 }{r_2 r_1 } + \frac{3}{l^2}(r_2 - r_1 ) + Q_0^2
\frac{r_2^3 - r_1^3 }{r_2^3 r_1^3 } = - \frac{1 - x}{r_2 x} + \frac{3r_2
}{l^2}(1 - x) + \frac{Q_0^2 }{r_2^3 x^3}(1 - x^3),
\end{equation}
\begin{equation}
\label{eq24}
8\pi T_0 = \frac{r_2 + r_1 }{r_2 r_1 } + \frac{3}{l^2}(r_2 + r_1 ) - Q_0^2
\frac{r_2^3 + r_1^3 }{r_2^3 r_1^3 } = \frac{1 + x}{r_2 x} + \frac{3}{l^2}r_2
(1 + x) - \frac{Q_0^2 }{r_2^3 x^3}(1 + x^3).
\end{equation}
Combining Eqs.(\ref{eq23}) and (\ref{eq24}), we can obtain
\begin{equation}
\label{eq25}
\frac{r_2^2 }{l^2}x = 1 - \frac{Q_0^2 (1 + x - 4x^2 + x^3 + x^4)}{r_2^2
x^2(1 - x)^2} = 1 - \frac{Q_0^2 (1 + 3x + x^2)}{r_2^2 x^2}.
\end{equation}
Substituting Eq.(\ref{eq25}) into Eq.(\ref{eq23}), we have
\begin{equation}
\label{eq26}
r_2^2 = \frac{Q_0^2 (1 + 4x + x^2)}{x^2} = Q^2f_Q (x).
\end{equation}
We find that Eq.(\ref{eq26}) is the same as Eq.(\ref{eq18}).
Substituting  Eqs.(\ref{eq25}) and (\ref{eq26}) into Eq.(\ref{eq24}), we can
obtain
\begin{equation}
\label{eq27}
2\pi T_0 xQf_Q^{3 / 2} (x) = (1 + x)f_Q (x) - \frac{(1 + 3x + 3x^2 +
x^3)}{x^2}.
\end{equation}
Taking $T_0 = \chi T_c$, where $T_c = \frac{\sqrt 6 }{18\pi Q}$ is the critical temperature. When $0<\chi<1$, the Eq.(\ref{eq27}) can be rewrite as
\begin{equation}
\label{eq28}
\chi x^2f_Q^{3 / 2} (x)\frac{1}{3\sqrt 6 } = (1 + x).
\end{equation}
From Eqs.(\ref{eq4}), (\ref{eq18}) and (\ref{eq19}), we can plot the $T - S$ phase
diagram with different $\chi$ in Fig. \ref{TSR}.

\subsection{The construction of equal-area law in~$P - v$~ diagram}
\label{subsubsec:mylabel3}

\begin{figure}[!htbp]
\center{
\includegraphics[width=7cm,keepaspectratio]{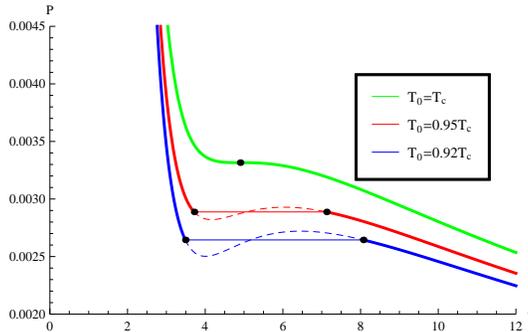}\hspace{0.5cm}}\\
\captionsetup{font={scriptsize}}
\caption{(Color online) The simulated phase transition and the boundary of a two-phase
coexistence on the base of isotherms in the $P-v$ diagram for charged anti-de Sitter black holes.
The temperature of isotherms from top to bottom. The green line is the critical isotherm diagram. we have set $Q=1$.
\label{Pv1R}}
\end{figure}

\begin{figure}[!htbp]
\center{
\includegraphics[width=7cm,keepaspectratio]{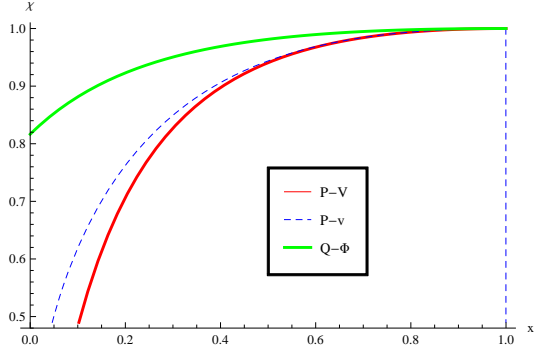}\hspace{0.5cm}}\\
\captionsetup{font={scriptsize}}
\captionsetup{font={scriptsize}}
\caption{(Color online) The $\chi-x$ diagram for charged anti-de Sitter black holes with differential duality state parameters.
The line corresponding with conjugate variables $P-V$, $P-v$ and $Q-\Phi$ from top to bottom.
\label{xx}}
\end{figure}

The horizontal axis of two-phase coexistence region in charged AdS black
hole are $v_2 = 2r_2 $ and $v_1 = 2r_1$, respectively, the pressure on the vertical axis as $P_0$.
From the Maxwell's equal area law,
\begin{equation}
\label{eq29}
P_0 (v_2 - v_1 ) = \int\limits_{v_1 }^{v_2 } {Pdv}
\end{equation}
So similarly, we can obtain
\begin{equation}
\label{eq30}
r_2^2 = Q^2\frac{4(1 - x^3) + 3(1 + x^2)(1 + x)\ln x}{x^23\left( {2(1 - x) +
(1 + x)\ln x} \right)} = Q^2F_Q (x).
\end{equation}
Taking $T_0 = \chi T_c$, here $T_c = \frac{\sqrt 6 }{18\pi Q}$ is critical temperature. When $0 < \chi <
1$, we have
\begin{equation}
\label{eq31}
\chi x^3F_Q^{3 / 2} (x)\frac{4}{3\sqrt 6 } = F_Q (x)x^2(1 + x) - (1 + x^2)(1
+ x).
\end{equation}
From Eqs. (\ref{eq30}) and (\ref{eq31}), we can plot phase diagram $P - v$ with different $\chi$ in Figs. \ref{Pv1R}.
From Eqs. (\ref{eq28}) and (\ref{eq31}), we can plot $\chi-x$ with differential duality conjugate variables in Fig. \ref{xx}.

From Eqs.(\ref{eq18}), (\ref{eq20}), (\ref{eq26}) and (\ref{eq28}), we know that the first-order
phase transition point of charged AdS black hole is same with differential conjugate variables $(P,V)$ and $(T,S)$ in same
temperature and the charge of AdS black hole is invariable. When we take the conjugate variable $(P,v)$, the first-order
phase transition point of charged AdS black hole is different from that of charged AdS black hole with
conjugate variable $(P,V)$ and $(T,S)$, however, the second-order phase transition is the same. We know that the first-order
phase transition point is independent of the the choice of  conjugate variable with the same temperature for the particle
numbers constant, so the conjugate variables $(P,v)$ can not be used as the states parameters of the
first-order phase transition of charged AdS black hole.

From Eqs.(\ref{eq18}),(\ref{eq26}) and (\ref{eq30}), the ratio of the phase transition point $r_2$ or $r_1$ to
the charge of black hole is constant with a fixed temperature. So the phase transition of charged
AdS black hole depends on the ratio of the phase transition point $r_2$ or $r_1$ to
the charge $Q$ of black hole, while it is independent of the radius $r_2$ or $r_1$ of the event horizon of
the black hole. When the charge $Q$ of charge AdS black hole is constant and the temperature is fixed,
the radius of the horizon of black hole $r_+<r_1$, the charged AdS black hole corresponds to the liquid phase of
van der Waals system, the radius of the horizon of black hole $r_ + > r_2 $, the charged AdS black hole
corresponds to the gas phase of van der Waals system, the radius of the horizon of black hole $r_2 \ge r_ + \ge r_1 $
corresponds to the gas-liquid coexistence of van der Waals system.
So when the temperature $T < T_c $ of the charged AdS black hole, there are three different states in the same pressure $P_0$.
From the different ratio of the charge to the radius of black hole horizon of black hole, We define these different
states as high potential state, medium potential state, low potential state.

\subsection{The construction of equal-area law in~$Q^2-\Psi$~and $Q-\Phi$ diagram}
\label{subsubsec:mylabel4}
\begin{figure}[!htbp]
\center{
\includegraphics[width=7cm,keepaspectratio]{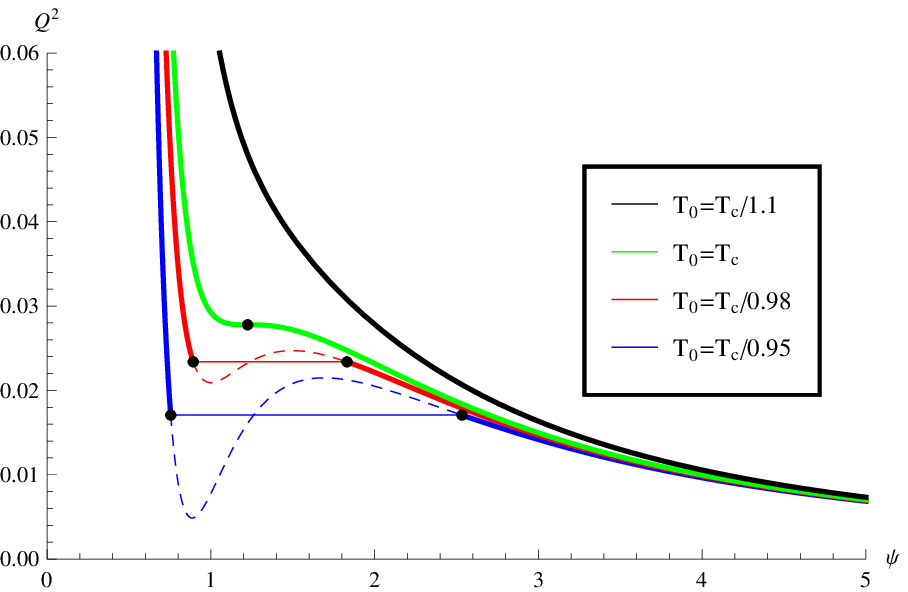}
\includegraphics[width=7cm,keepaspectratio]{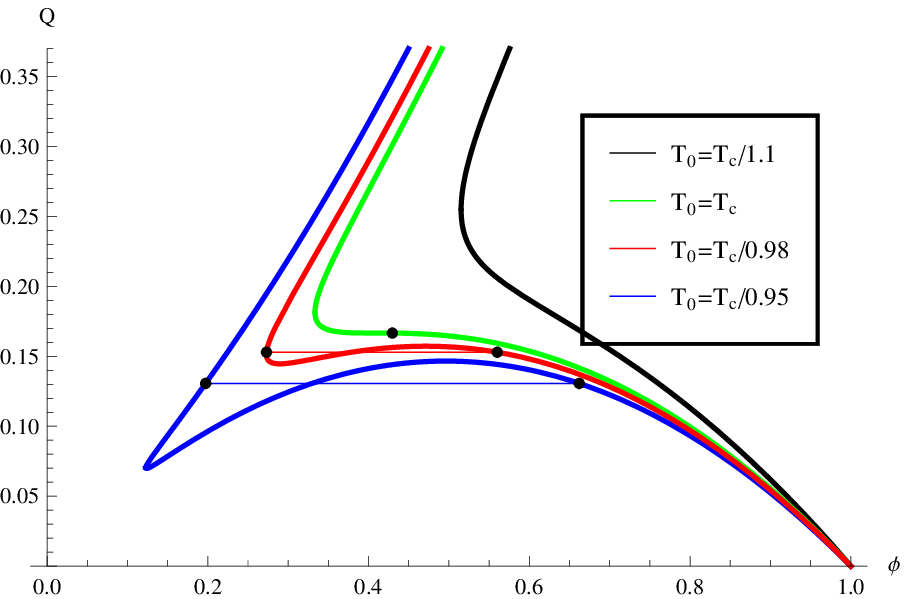}\hspace{0.5cm}}\\
\captionsetup{font={scriptsize}}
\caption{(Color online)(left) The behavior of isothermal $Q^2-\Psi$ diagram of charged anti-de Sitter black holes.
(right) The behavior of isothermal $Q-\Phi$ diagram of charged anti-de Sitter black holes.
The temperature of isotherms increases from top to bottom. The grren line is the critical isotherm diagram. we have set $l=1$.
\label{Q1}}
\end{figure}

When the charge of the AdS black hole of space is variant, taking
$Q^2-\Psi$ as independent duality parameter. The horizontal axis of
two-phase coexistence region in charged AdS black
hole are $\Psi_2$ and $\Psi_1$, respectively, the temperature on the vertical
axis as $Q_0^2$, which is dependent on the horizon radius of black hole $r_+$.
From the Maxwell's equal area law,
\begin{equation}
\label{eq32}
Q_0^2 (\Psi _1 - \Psi _2 ) = \int\limits_{\Psi _2 }^{\Psi _1 } {Q^2d\Psi }
.
\end{equation}
\begin{equation}
\label{eq33}
Q_0^2 = r_2^2 + \frac{3r_2^4 }{l^2} - 4\pi r_2^3 T_0 ,
\quad
Q_0^2 = r_1^2 + \frac{3r_1^4 }{l^2} - 4\pi r_1^3 T_0.
\end{equation}
we can obtain
\begin{equation}
\label{eq34}
Q_0^2 = r_2^2 x + \frac{r_2^4 }{l^2}x(1 + x + x^2) - 2\pi T_0 r_2^3 x(1 +
x),
\end{equation}
and
\begin{equation}
\label{eq35}
0 = (1 + x) + \frac{3r_2^2 }{l^2}(1 + x^2)(1 + x) - 4\pi T_0 r_2 (1 + x +
x^2),
\end{equation}
\begin{equation}
\label{eq36}
2Q_0^2 = r_2^2 (1 + x^2) + \frac{3r_2^4 }{l^2}(1 + x^4) - 4\pi T_0 r_2^3 (1
+ x^3).
\end{equation}
Combining Eqs. (\ref{eq34}), (\ref{eq35}) and (\ref{eq36}), we can get
\begin{equation}
\label{eq37}
r_2 = \frac{(1 + x)}{\pi T_0 (1 + 4x + x^2)} = \frac{f_l (x)}{\pi T_0 }.
\end{equation}
Taking $T_0 = \frac{T_c }{\chi } = \frac{1}{\chi \pi l}\sqrt
{\frac{2}{3}} $, we can obtain~\cite{Dehyadegari}
\begin{equation}
\label{eq38}
3\chi ^2(1 + x)^2 = 2(1 + 4x + x^2).
\end{equation}
Taking $(Q,\Phi)$ as independent conjugate variables, The horizontal axis of
two-phase coexistence region in charged AdS black
hole are $\Phi_2$ and $\Phi_1$, respectively, the pressure on the vertical
axis as $Q_0$. From the Maxwell's equal area law,
\begin{equation}
\label{eq39}
Q_0 (\Phi _1 - \Phi _2 ) = \int\limits_{\Phi _2 }^{\Phi _1 } {Qd\Phi } .
\end{equation}
So similarly, we can obtain
\begin{equation}
\label{eq40}
r_2 = \frac{(1 + x)}{\pi T_0 (1 + 4x + x^2)} = \frac{f_l (x)}{\pi T_0 }.
\end{equation}
Seting $T_0 = \frac{T_c }{\chi } = \frac{1}{\chi \pi l}\sqrt {\frac{2}{3}}$, we can obtain
\begin{equation}
\label{eq41}
r_2 = \frac{(1 + x)}{\pi T_0 (1 + 4x + x^2)} = \chi l\sqrt {\frac{3}{2}} f_l(x)£¬
\end{equation}
\begin{equation}
\label{eq42}
3\chi ^2(1 + x)^2 = 2(1 + 4x + x^2).
\end{equation}
From Eqs. (\ref{eq4}), (\ref{eq40}) and (\ref{eq42}), we can plot the phase diagram
$Q^2-\Psi $ and $Q-\Phi$ with different $\chi$ in Fig. \ref{Q1}. We can plot the $\chi - x$
from (\ref{eq41}) in Fig.\ref{xx}.
From Eqs.(\ref{eq37}), (\ref{eq38}), (\ref{eq40}) and (\ref{eq42}), we find that the phase transition point is same for the two conjugate variables ($Q^2, \Psi)$ and ($Q,\Phi )$ with the fixed temperature. So, the first laws of thermodynamics represented by the two conjugate variables ($Q^2,\Psi )$ and($Q, \Phi )$ are equivalent.

From Eq.(\ref{eq41}), the ratio of the phase transition point $r_2$ or $r_1$ to
the cosmological constant $l$ of space is constant with a fixed temperature for the particle number variance.
So the phase transition of charged AdS black hole dependent with the ratio of the phase transition point $r_2$ or $r_1$ to
the cosmological constant $l$, is independent of the radius $r_2$ or $r_1$ of the event horizon of
the black hole. When the cosmological constant $l$ is constant and the temperature $T > T_c$ is fixed,
the radius of the horizon of black hole $r_+<r_1$, the charged AdS black hole corresponds to the liquid phase of
van der Waals system, the radius of the horizon of black hole $r_ + > r_2 $, the charged AdS black hole
corresponds to the gas phase of van der Waals system, the radius of the horizon of black hole $r_2 \ge r_ + \ge r_1 $
corresponds to the gas-liquid coexistence of van der Waals system.
So when the temperature $T > T_c $ of the charged AdS black hole, there are three different states in the same pressure $Q$.
From the different ratio of the radius of black hole horizon of black hole to the cosmological constant $l$ of space,
We define these different states as larger phase, medium phase, small phase.

\section{Discussions and conclusions}
\label{con}
The study of black hole thermodynamics have gained attention as the black hole, especially the thermodynamics
of black hole directly involve the gravitation,  statistics, particle physics and field theory~\cite{Bekenstein,Bekenstein2,Bekenstein3,Bardeen4,Hawking,Hawking2}. Even so it is unclear in the
precise statistical description of a black hole's thermodynamic state the thermodynamic
properties of black holes and the critical phenomenon in ordinary thermodynamic systems still is a research
subject widely concerned.

In this paper, we choose different conjugate variables and use Maxwell's equal-area law to
study the phase transition of charged AdS black hole. We find that the phase transition point
which obtained by choosing the conjugate variable $(P,V)$ and $(T,S)$, respectively, is equal
at the same temperature. The phase transition point obtained by the conjugate variable $(P,v)$ is different
with that obtained by choosing the conjugate variable $(P,V)$ and $(T,S)$. For the phase transition point of
AdS black hole is the states function of system, and is independent of the process, so we think that
the conjugate variable $(P,v)$ cannot be independent conjugate variable to study the phase transition of AdS black
hole.

From the study of these section \ref{subsubsec:mylabel1}, \ref{subsubsec:mylabel2} and \ref{subsubsec:mylabel3}, we obtain that
when the temperature $T < T_c $ of the charged AdS black hole, there are three differential states in the same pressure $P_0$.
From the differential ratio of the charge to the radius of black hole horizon of black hole, We define these different
states as high potential state, medium potential state, low potential state.From the study of this section
\ref{subsubsec:mylabel4}, when the temperature $T > T_c $ of the charged AdS black hole, there are three
differential states in the same pressure $Q$. From the differential ratio of the radius of black hole horizon
of black hole to the cosmological constant $l$, We define these different states as larger phase,
medium phase, small phase. As we have seen, the phase transition of charged AdS black hole, not only depend on
the event horizon radius of black hole, but also depend on the ratio the charge potential to the radius of black hole
horizon. It would be worthwhile to
better understand the critical behaviour of black hole, and perfect the self-consistent geometry theory of black hole thermodynamic.

\section*{Acknowledgements}
We would like to thank Prof. Zong-Hong Zhu and Meng-Sen Ma for their indispensable discussions and comments. This work was supported by the Young Scientists Fund of the National Natural Science Foundation of China (Grant No.11205097), in part by the National Natural Science Foundation of China (Grant No.11475108), Supported by Program for the Innovative Talents of Higher Learning Institutions of Shanxi, the Natural Science Foundation of Shanxi Province,China(Grant No.201601D102004) and the Natural Science Foundation for Young Scientists of Shanxi Province,China (Grant No.2012021003-4).

\end{document}